\documentclass[useAMS,usenatbib]{mn2e}
\usepackage{amsmath,amsfonts,epsfig,natbib}

\def\reff@jnl#1{{\rm#1\/}}

\def\aj{\reff@jnl{AJ}}                  
\def\araa{\reff@jnl{ARA\&A}}            
\def\apj{\reff@jnl{ApJ}}                        
\def\apjl{\reff@jnl{ApJ}}               
\def\apjs{\reff@jnl{ApJS}}              
\def\ao{\reff@jnl{Appl.Optics}}         
\def\apss{\reff@jnl{Ap\&SS}}            
\def\aap{\reff@jnl{A\&A}}               
\def\aapr{\reff@jnl{A\&A~Rev.}}         
\def\aaps{\reff@jnl{A\&AS}}             
\def\azh{\reff@jnl{AZh}}                        
\def\baas{\reff@jnl{BAAS}}              
\def\jrasc{\reff@jnl{JRASC}}            
\def\memras{\reff@jnl{MmRAS}}           
\def\mnras{\reff@jnl{MNRAS}}            
\def\pra{\reff@jnl{Phys.Rev.A}}         
\def\prb{\reff@jnl{Phys.Rev.B}}         
\def\prc{\reff@jnl{Phys.Rev.C}}         
\def\prd{\reff@jnl{Phys.Rev.D}}         
\def\prl{\reff@jnl{Phys.Rev.Lett}}      
\def\pasp{\reff@jnl{PASP}}              
\def\pasj{\reff@jnl{PASJ}}              
\def\qjras{\reff@jnl{QJRAS}}            
\def\skytel{\reff@jnl{S\&T}}            
\def\solphys{\reff@jnl{Solar~Phys.}}    
\def\sovast{\reff@jnl{Soviet~Ast.}}     
\def\ssr{\reff@jnl{Space~Sci.Rev.}}     
\def\zap{\reff@jnl{ZAp}}                        
\def\nat{\reff@jnl{Nature}}             

\title[VSA first results III -- power spectrum]{First results from the VSA --  
III. The CMB power spectrum}

\author[Paul F. Scott et al.]
{Paul F. Scott$^1$, 
 Pedro Carreira$^2$, Kieran Cleary$^2$, Rod D. Davies$^2$, Richard J.
 Davis$^2$,  \newauthor Clive Dickinson$^2$, Keith Grainge$^1$, Carlos M.
 Guti{\'e}rrez$^3$,  
 Michael P. Hobson$^1$,  \newauthor
 Michael E. Jones$^1$, R\"udiger Kneissl$^1$,
 Anthony Lasenby$^1$,  
 Klaus Maisinger$^1$,  \newauthor
 Guy G. Pooley$^1$, Rafael Rebolo$^{3,4}$, Jos\'e Alberto
 Rubi\~no-Martin$^3$,  
 Pedro Sosa Molina$^3$,  \newauthor Ben Rusholme$^{1,\star}$, Richard
 D.E. Saunders$^1$, Richard Savage$^1$,  
 An\v ze Slosar$^1$, \newauthor Angela C. Taylor$^1$,  David
 Titterington$^1$,  
 Elizabeth Waldram$^1$, \newauthor Robert A. Watson$^{2,\dagger}$, Althea Wilkinson$^2$
\\
  $^1$ Astrophysics Group, Cavendish Laboratory, University of Cambridge, UK\\
  $^2$ University of Manchester, Jodrell Bank Observatory, UK\\
  $^3$ Instituto de Astrof{\'i}sica de Canarias, 38200 La Laguna,
  Tenerife, Spain.\\
  $^4$Consejo Superior de Investigaciones Cient{\'{\i}}ficas, Spain \\
  $^{\star}$Present address: Stanford University, Palo Alto, CA, USA\\
  $^{\dagger}$Present address: Instituto de Astrof{\'{\i}}sica de
Canarias.}

  \pagerange{\pageref{firstpage}--\pageref{lastpage}}

\pubyear{2002}

\begin{document}
\maketitle
\label{firstpage}
\begin{abstract}
We present the power spectrum of the fluctuations in the cosmic microwave
background detected by the Very Small Array (VSA) in its first season of
observations in its compact configuration. We find clear detections of first
and second acoustic peaks at $\ell{\approx 200}$ and $\ell{\approx 550}$, plus
detection of power on scales up to $\ell=800$.  The VSA power spectrum is in
very good agreement with the results of the BOOMERANG, DASI and MAXIMA
telescopes despite the differing potential systematic errors.
\end{abstract}


\section{Introduction}

Anisotropies in the Cosmic Microwave Background (CMB) have now been detected
by many experiments (most recently \citet{netterfield-01}, \citet{lee-01},
\citet{halverson-02}, \citet{padin-01}). At present the most successful model
for explaining the origin of these fluctuations postulates that they are
seeded in the very early universe by quantum perturbations with random phase,
followed by a period of inflationary expansion.  The fluctuations in the CMB
are predicted to be Gaussian in nature, and hence can be completely
characterised through their power spectrum. A further prediction is that the
power spectrum will show acoustic peaks due to plasma oscillations on scales
smaller than the sound horizon at the surface of last scattering.

In this paper we present the power spectrum of the CMB fluctuations detected
by the Very Small Array (VSA) on spherical harmonic modes $\ell
\approx$150--900. We outline how the fully calibrated time-stream data are
converted into a power spectrum and the various data checks that we have
performed to confirm the validity of our analysis. This paper is the third in
a series of four papers which report the results of the first season of
observations made using the VSA in its compact configuration.
Paper~I~\citep{VSApaperI} describes the design of the VSA and our experimental
method; the observational strategy, foreground removal and reduction
techniques for the data analysed in this paper are described in
Paper~II~\citep{VSApaperII}; finally, the cosmological implications of the VSA
power spectrum are discussed in \citet{VSApaperIV} (Paper IV).

\section{Observations and Initial Data Processing}

\subsection{The observations}

The Very Small Array (VSA) is a 14-element interferometer array for Cosmic
Microwave Background (CMB) observations; it operates at a frequency between 26
and 36 GHz with a receiver bandwidth of 1.5 GHz.  In its compact
configuration, used here, the instrument is most sensitive to angular
structure in the range $\ell \approx$150--900. As well as the 14-element
array, there is also a single baseline interferometer used for radio source
flux measurements. A fuller description of the instrument is given in Paper I.

The present observations were made at a frequency of 34~GHz in the period 2000
September to 2001 September and were centred on three separate areas of sky.
Each VSA observation maps a region with FWHM of $4.6^\circ$.  The process of
field selection is discussed in Paper II; selection was based primarily on low
Galactic (synchrotron, free-free and dust) emission and an absence of known
bright foreground sources.  Overlapping fields were observed in each area in
order to reduce the sample variance, to increase the resolution in
$\ell$-space and to allow for direct assessment of data reliability and the
detection of any residual instrumental effects (Section~\ref{spurious}). The
array configuration used was designed to provide an approximately uniform
spread of interferometer baselines while retaining a reasonable aperture
filling factor.

The amplitude and phase calibrations of the individual interferometer
baselines were checked both by frequent short measurements of a number of
secondary calibration sources and also by regular longer observations of a
small number of primary calibration sources.  The overall calibration
procedure is described fully in Paper I; the overall accuracy of the
calibration in flux density and temperature is $3.5\%$ .

An important aspect of the VSA is the inclusion of a separate interferometer,
comprising two 3.7-m dishes on a 9-m north-south baseline, for determining the
flux densities of foreground sources (radiogalaxies and quasars) falling
within the observed fields.  The positions of all sources which might affect
the VSA observations were obtained from survey observations using the Ryle
Telescope ~\citep{waldram-02}; these positions were subsequently observed
concurrently with the main VSA observations in a series of regular pointed
observations and the measured flux densities subtracted from the visibilities
observed by the main array (Paper II).

\subsection{Initial Data Processing}

The data are calibrated and processed as described in Papers I and II. Early
tests of the telescope revealed an unwanted local spurious signal,
particularly evident on the shorter baselines.  The procedures for removing
this signal from the data are described fully in Paper II; the tests presented
in Section~\ref{spurious} below indicate that the removal of this signal is
complete.  Similar filtering processes were used to eliminate signals arising
from the Sun and the Moon.

\section{Derivation of the Power Spectrum}\label{derivation}

Since the three sets of mosaiced fields are widely separated from one
another
on the sky, the data corresponding to each field are analysed
individually. The derivation of the CMB power spectrum for each mosaic is
performed using the maximum-likelihood method presented in \citet{klaus},
which we summarise here. The results from each mosaic are then combined,
as
outlined below, to produce the final estimate of the power spectrum.

Since the number of individual visibility measurements for each mosaic is
very
large ($\sim$ 800,000), it is first necessary to compress these data in
some
way. For each separate field in the mosaic, the visibilities are binned
into
cells in the $uv$-plane; for the case in which the instrumental noise
covariance
matrix is diagonal, this corresponds to the maximum-likelihood
solution for the value of the binned signal visibility in each cell.
This is
analogous to the `map-making' step in the analysis of single-dish CMB
observations, in which time-ordered data are binned into pixels on the sky
(see, e.g. \citet{Borrill99}). Since we are not interested here in making
accurate CMB maps from the binned visibility data, the $uv$-plane is
simply
divided into equal-area cells of size $\Delta u= 3$ wavelengths. As the
aperture function of the compact VSA is well-modelled by a Gaussian with a
FWHM of 12 wavelengths, this cell size ensures that the $uv$-plane is
comfortably over-sampled, while reducing the total number of visibility
measurements for each field significantly (to~$\sim$ 2500).

The binned visibilities are thus the basic input to the likelihood
analysis
for the CMB power spectrum.  The compact VSA is sensitive to the multipole
range $\ell\sim150-900$, and the effective aperture function (after
mosaicing)
has a FWHM of $\Delta \ell = 83$. We thus divide the total $\ell$-range
into
10 spectral bands each of width $\Delta\ell$, in order to limit the
correlation between adjacent bands. In each band, we assume
$\ell(\ell+1)C_\ell$ to have a constant value $\bar{\cal C}_B$
$(B=1,2,\ldots,10)$. These flat power levels in each band are then the
parameters to be determined in our likelihood analysis of the data for
each
mosaic. We denote these band powers collectively by the parameter vector
$\bmath{a}$.

Assuming the CMB emission and the instrumental noise to be Gaussian random
fields, the log-likelihood of obtaining the binned visibility data vector
$\bmath{v}$ (which contains the real and imaginary parts separately of
each
binned visibility), given some set of flat band powers $\bmath{a}$, is
given
by
\[
\ln{\cal L} = \mbox{constant}-{\textstyle\frac{1}{2}}
\left[\ln|\bmath{C}(\bmath{a})|+\bmath{v}^{\rm
t}\bmath{C}^{-1}(\bmath{a})\bmath{v}\right],
\label{loglike}
\]
where $\bmath{C}(\bmath{a})=\bmath{S}(\bmath{a})+\bmath{N}$ is the sum of
the
predicted signal covariance matrix and the noise covariance matrix. The
maximum-likelihood CMB power spectrum $\hat{\bmath{a}}$ is calculated
using a
simple iterative numerical maximisation algorithm. Starting from an
initial
guess $\bmath{a}_0$ (which is unimportant), independent line maximisations
are
performed for each band power ${\cal C}_B$ in turn, while keeping the
others
fixed. The whole solution vector $\bmath{a}$ is then updated and the
process
repeated until convergence is obtained. This typically requires around 5
iterations.

The well-defined correlation structure of visibility data in the
$uv$-plane
allows each line-maximisation to be performed using only the subset of
visibilities that are sensitive to the band power being varied, thereby
speeding-up the evaluation of the likelihood function, while keeping
the calculation exact. For a single VSA
mosaic, the maximum-likelihood solution can be obtained in around one hour
on an 8-node Beowulf Cluster with 1.8-GHz AMD Athlon processors.

The uncertainties in the derived maximum-likelihood CMB power spectrum are
estimated in two complementary ways. Assuming the likelihood function in
the
parameter space $\bmath{a}$ to be well-approximated by a multivariate
Gaussian
near its peak, the covariance matrix of the parameter uncertainties is
given
simply by (minus) the inverse of the curvature (or Hessian) matrix at the
peak
$\hat{\bmath{a}}$. This matrix is easily evaluated numerically in
a few hours of CPU time.
The square-root of the diagonal elements of the resulting
covariance matrix give the standard error $\Delta \bar{\cal C}_{\rm B}$ on
each band power, whereas the off-diagonal elements provide a measure of
the
correlation between the band power estimates in different spectral bins.
We
find, typically, that adjacent bins are anti-correlated at around the
$5-20\%$
level, and for more widely separated bins the correlation is negligible.

Using the covariance matrix clearly produces symmetric error bars on each
band
power $\bar{\cal C}_{\rm B}$, which may be a poor representation of the
uncertainty, especially for poorly constrained band powers.  An
alternative
approach is to make use of the fact that the band power estimates in
different
spectral bins are only weakly correlated, so that the off-diagonal
elements of the
covariance matrix are small compared to those on the diagonal.  In this
case,
a better representation of the uncertainty in each estimate may be
obtained by
directly evaluating the likelihood function through the peak
$\hat{\bmath{a}}$, along each parameter direction in turn. In the ideal
case,
where the band power estimates are independent, the resulting curves would
be
the marginal distributions of each band power. Since the likelihood
function
can be evaluated very quickly along each direction in parameter space, the
resulting `marginal' distributions can be calculated for each VSA mosaic
in
around 2 hours, and provide a useful cross-check of the standard errors
obtained from the covariance matrix at the peak. We find that for
spectral bins in which the flat band power is tightly-constrained, the
shape of the likelihood function is very close to Gaussian. However,
for bins in which the power level less well constrained, the
likelihood function is better described by an offset log-normal
distribution; this is discussed further in Paper IV.

Once the likelihood functions for the flat band power $\bar{\cal C}_{\rm
B}$
in each spectral bin have been obtained separately for each VSA mosaic,
the
mosaics are combined simply by multiplying together the respective
likelihood
functions in each bin.  This assumes that each mosaic provides an
independent
measurement of the CMB power spectrum in each spectral bin, which is valid
given that the three sets of mosaiced fields are widely separated on the
sky.
The correlation between the resulting band power estimates in different
bins
is easily obtained by calculating the covariance matrix, as described
above,
at the new joint optimum $\hat{\bmath{a}}$.

All the above functions are implemented using the {\sc Madcow} analysis
package \citep{klaus}.

\subsection{Window functions}

Because of various instrumental effects (e.g. non-uniform $uv$ coverage and
the finite size of the primary beam of the telescope), a bin samples the
underlying power spectrum $C_l$ through a window function $W(l)$.
For the \textit{B}th bin, the measured power corresponds to

\[
p_B =\sum_\ell \frac{W_B(\ell)}{\ell} {\cal C}_\ell
\]
where ${\cal C}_\ell = \ell(\ell+1)C_\ell/(2\pi)$.

The window function for a given bin is determined as follows. The 
range of Fourier modes
to which a given visibility is sensitive are given by the Fourier transform
of the primary beam
for a single field observation (see e.g. \citet{klaus}).
For a Gaussian primary beam,

\[ 
\tilde{A}(\textbf{u})=2\pi\sigma^2\exp(-2\pi^2\sigma^2|\textbf{u}|^2),
\]
where $\textbf{u}$ is the vector in the visibility plane (measured in
wavelengths). For mosaiced observations, the effective beam is a superposition
of displaced primary beams. We can think of this superposition as a
convolution of a centred primary beam with a sum of delta
functions at the beam centres. In the Fourier domain this corresponds to a multiplication of the
aperture function with the `Fraunhofer diffraction pattern' of the beam
centres

\[
\tilde{A}_{\mbox{eff}}(\textbf{u})=\Sigma_{j}\tilde{A}(\textbf{u})e^{2\pi i \textbf{u} \cdot \textbf{x}_j},
\]
where $\textbf{x}_j$ are the positions of the beam centres from the origin,
measured in radians. Note that in general this aperture function is complex.

For a given bin, the weighted complex sensitivity map is
\[
S(\textbf{u})=\sum_k w_k \tilde{A}_{\mbox{eff}} (\textbf{u}-\textbf{v}_k),
\]
where index $k$ runs over visibilities in a given bin. These visibilities have
instrumental weights $w_k$ and positions on the visibility plane
$\textbf{v}_k$. 

\begin{figure}
        \epsfig{file=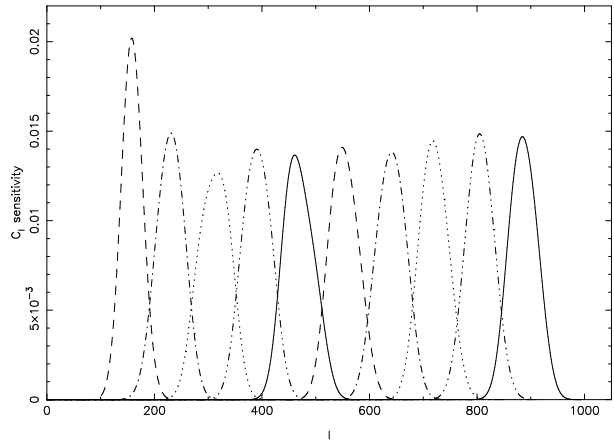,angle=-90,width=8cm}
\caption{Window functions for the combined data set. The functions are
normalised to unit area, and different bins are plotted with different
linestyles to allow easier visual differentiation. \label{fig:win}}
\end{figure}

The un-normalised window function is then given by:
\[
W(\ell) = \mathcal{N} \int_{\phi=0}^{\phi=2\pi} |S(\ell,\phi)|^2 \, {\rm
d}\phi ,
\]
where $\ell=2\pi|\textbf{u}|$. The normalisation constant
$\mathcal{N}$ can be trivially found using $\sum_\ell (2\ell+1)
W(\ell)/(2\ell(\ell+1))=1$.  The window functions for the combined VSA
data set are plotted in Figure~\ref{fig:win}, showing the small degree
of correlation between adjacent bins.  These functions are used in
Paper IV in the estimation of cosmological parameters.

\section{Data Checks}

The complete process of editing and filtering the data and also the subsequent
stages of data reduction were carried out independently by the Cambridge group
and, jointly, by the combined IAC and JBO teams.  A comparison of the two sets
of results showed good agreement, the effect of any differences being small
compared to the intrinsic uncertainties on the final power spectra.

\subsection{Test of data reduction procedure}

Aspects of our data reduction procedure, such as filtering and calibration,
could potentially have introduced systematic errors into the VSA data. In
order to test this, we produced a realisation of the CMB sky and simulated a
mock VSA observation including such instrumental effects as visibility
quadrature errors, phase steps due to path compensation and thermal noise. We
analysed these simulated data using our standard reduction procedure and then
produced a power spectrum as described in section~\ref{derivation}. We found
this to be entirely consistent with our input model and so concluded that our
method of data reduction did not introduce any significant systematic errors.
In particular, it demonstrates that any correlations in the visibilities 
introduced by the filtering process has not observable effect on the
derived power spectrum.

\subsection{The Local Spurious Signal}\label{spurious}

The local spurious signal, fully discussed in Paper~I, was found to depend
only on the antenna tracking angle, and not on the table elevation.
Therefore, we would expect the spurious signal to be identical for different
fields with the same declination observed over the same hour angle range.
This is easily confirmed by combining the (unfiltered) baseline timeseries of
two such fields by both addition and subtraction.  We find that adding the two
fields enhances the spurious signal, whilst subtraction entirely removes it.
Whilst this technique of addition and subtraction is adequate for detecting
the presence of spurious signal in unfiltered data, it is insufficiently
sensitive to detect possible low level residual signal once filtering has been
applied.

The increase in detection sensitivity that we require in order to test the
filtered data for residual spurious signal can, however, be obtained using a
modified version of a MEM algorithm used for extraction of CMB signal from VSA
data~\citep{MHL97}.

We add an extra term to the MEM reconstruction which is the signal that is
identical in the CMB datasets. We then consider the case of two fields at
identical declination.  As the CMB signal in the two fields will not be
identical, and the noise is random in each case separately, any common
component to the two fields will be spurious signal.

\begin{figure}
\centering  
\epsfig {file=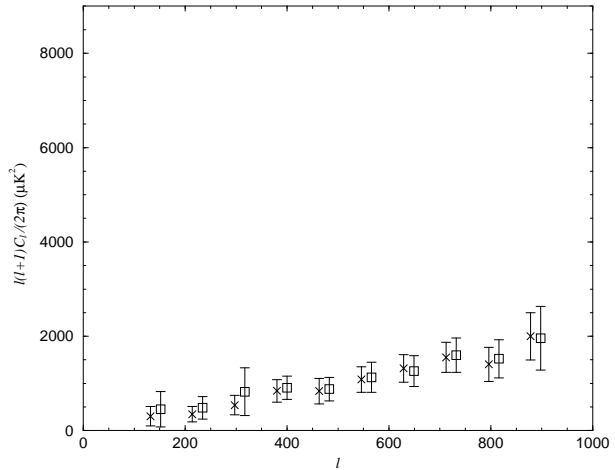,height=8cm,angle=270} 
\caption{Power spectra of the recovered common signal from the two VSA2
fields (squares) and from a 
pair of simulated datasets with no common signal (i.e. independent noise
only) (crosses).  The pairs of points have been separated laterally for clarity.
Coverage of the $uv$-plane and the
thermal noise level are identical for both the real data and the simulation.
\label{maxspur}}
\end{figure}

We tested this algorithm by applying it to pairs of simulated observations of
CMB fields assuming CDM primordial fluctuations, to which we added an
identical component with an rms level such that it was not the dominant
signal.  We used a variety of common components, including a scaled down
version of the unfiltered spurious signal. We found that the MEM algorithm was
able to reconstruct the common signals well, recovering the structure
excellently, and recovering the amplitude of the signal to within 10\%.  As we
are primarily interested in whether or not the spurious signal is still
present, as opposed to any accurate quantification, this is perfectly
adequate.  Note that, even for entirely independent fields, the reconstructed
shared signal power spectra are not zero (Figure~\ref{maxspur}),
but rather show a value increasing as $\ell^2$,  consistent with the
correlation between two white noise signals.

The two VSA2 fields had identical declination to allow testing of our
filtering procedures.  The MEM algorithm was applied to these fields, and the
results compared with CDM realisations with identical uv-coverage and thermal
noise in which no shared signal is present. The results (Figure~\ref{maxspur})
show that the common signal found from the real data fields and the
simulations agree within the errors, the difference between the two
sets of points being less than 4\% of the measured CMB for values of
 $\ell < 750$.    In the same manner,
we find no evidence for residual spurious signal in the pairs of VSA1,3 fields
at similar declinations.

Non-Gaussianity analysis of the binned visibilities allows us to locate and
remove the few remaining visibilities contaminated with spurious signals down
to a low level.  The removal of these points has a negligible effect on the
final power spectrum, giving us confidence that we are subtracting the
spurious signal to a level well below that which could affect our results. The
full details of this analysis will be published in Savage et al. in prep.

\subsection{Foreground Source Subtraction}

\begin{figure}
 \centering
   \epsfig {file=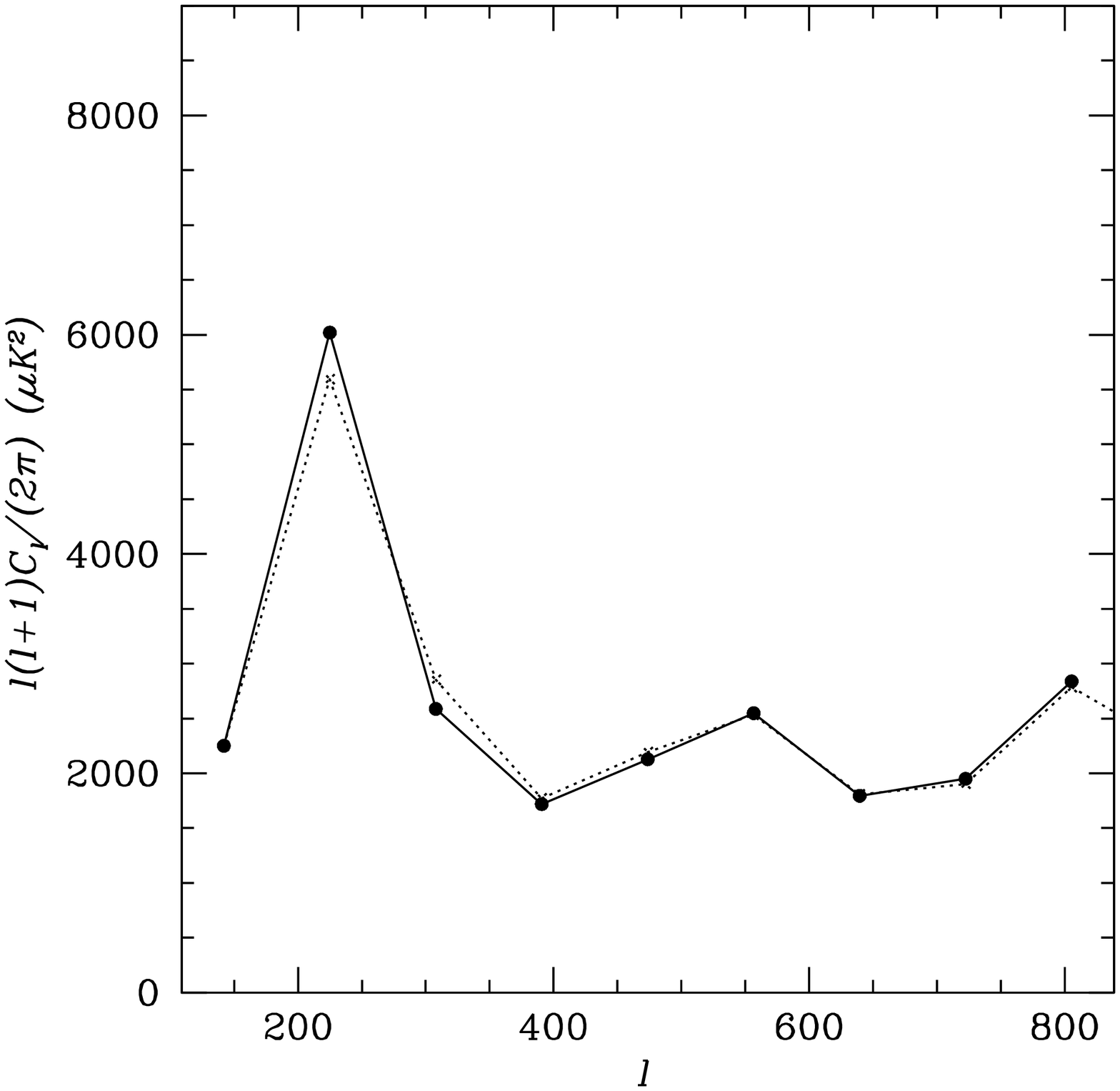,height=6cm}
   \epsfig{file=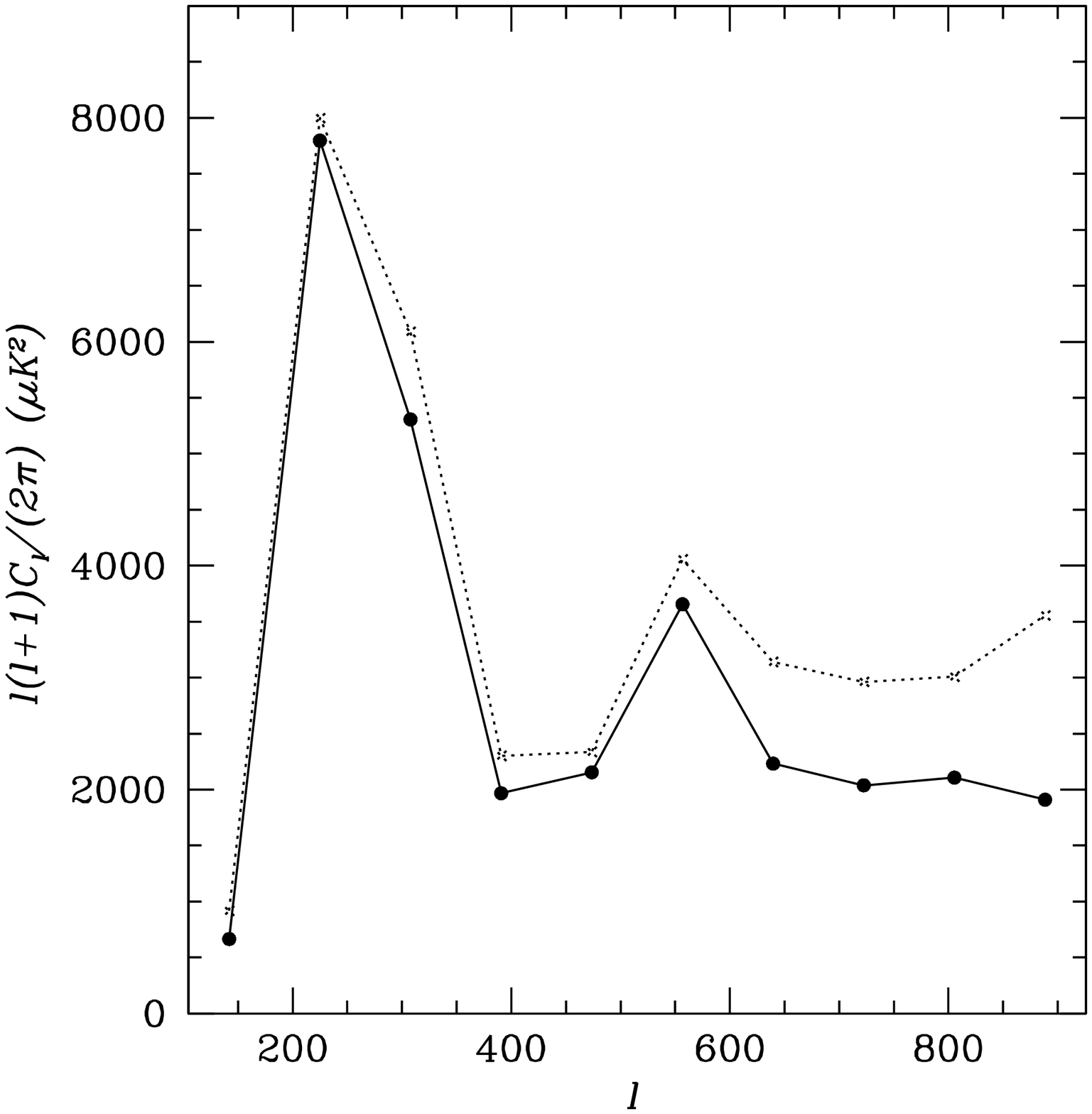,height=6cm}
\caption{Two examples of recovered power spectra from simulated CMB
observations with (filled circles, solid lines) and without (open circles,
dotted lines) the known sources in the VSA1 field added. The differences arise
from chance interactions between the sources and individual CMB features.}
\label{CMBsource_sim}
\end{figure}

Radio galaxies and quasars are a significant contaminant of the CMB at
microwave frequencies and in the higher $\ell$-ranges will dominate the CMB
signal, making it essential to remove their contribution.  Tests with
simulated fields have been used to assess the potential contribution, before
subtraction, of these sources to the final CMB spectrum. We generated ten
realisations of the CMB sky using a particular CDM model and added to these
the sources that we have observed to be present in the VSA1 field. We have
then compared the power spectra recovered from simulated VSA observations with
and without sources.  Two of these simulations, representing the range of
results obtained from the ten simulations, are shown in
Figure~\ref{CMBsource_sim}.  It is apparent that, with the fairly small number
of sources in any one VSA field (typically 12 sources), the impact on the
power spectrum is unpredictable.  Although the main contribution tends to be
in the higher $\ell$-bins (with errors potentially reaching $\pm$100\% at
$\ell=900$), changes of up to $\pm$10\% can occur in the lowest $\ell$-bin;
these are due to the chance superposition of interference fringes.

\begin{figure}
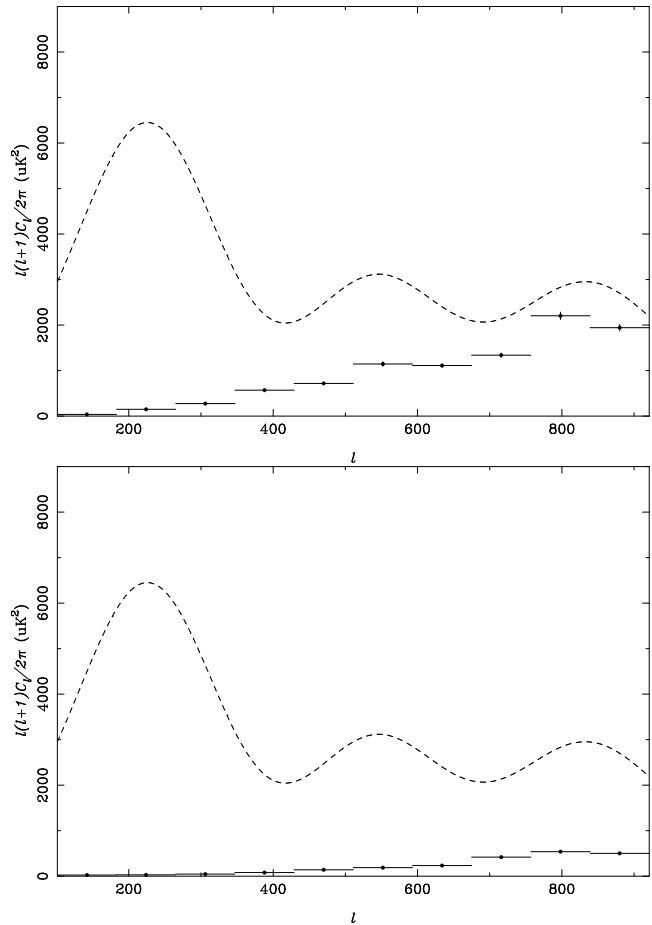

 \centering
   \epsfig{file=sources.ps,height=85mm,angle=-90}
   \epsfig{file=sub_sources.ps,height=85mm,angle=-90}
 \caption{Simulated power spectra for known sources plus a
statistical distribution of weaker sources (upper plot) and
for the distribution of weaker sources alone (lower plot). For
comparison the dashed curve shows the predicted power spectrum for a
CDM model.}
 \label{sources}
\end{figure}

To determine the effect of residual sources on the VSA results, two further
simulations {\it without} a CMB contribution have been carried out.  We base
these simulations on the 15-GHz source counts from \citet{TGJ01} and
extrapolate up to 34~GHz using a mean spectral index of $\alpha=0.55$.  The
first (Figure~\ref{sources} upper plot) comprises the contributions of the
known point sources {\it and} a statistical distribution of weaker sources;
the second (lower plot) includes only the statistical contribution of weaker
sources and gives an indication of the possible residual contribution to the
VSA power spectrum {\it after} subtraction of known sources.  It is clear
that, with no source subtraction, the CMB data are significantly compromised
for $\ell$-values $\ge 600$.  After subtraction of known sources, the
contribution is reasonably small for $\ell$-values up to about 1000.  As
demonstrated by the results of Figure \ref{CMBsource_sim}, the contribution of
sources to the observed spectra can not generally be predicted from a simple
combination of simulated power spectra, although such an approach becomes
feasible in the limiting case of many weak sources per synthesised beam.

We can also estimate the residual source contribution to the power spectrum
using the preliminary 34~GHz source count derived from the source subtractor
observations in Paper II. Integrating the count from zero flux to our complete
source subtraction limit of 80~mJy and converting to units of $\Delta T/T_0$,
the source power is given by $C_{\ell} = 7.7 \times 10^{-16}$, corresponding
to a power spectrum value of $\frac{\ell(\ell + 1)}{2 \pi}C_{\ell}  = 580
\, \mu$K$^2$ at $\ell = 800$. For the case with effectively no source
subtraction, taking the upper flux limit to be $0.5$~Jy, we find $C_{\ell} =
3.7 \times 10^{-15}$, equivalent to $2800 \, \mu$K$^2$ at $\ell = 1000$;  these
results are in good agreement with the extrapolations from the 15 GHz
counts.  

The precise residual contribution due to  sources is subject to uncertainties in 
our knowledge of the true weak-source distribution;  the values shown in
Figure~\ref{sources} are likely to be  over-estimates since some sources with
flux densities less than 80~mJy were actually subtracted.   In order to assess the possible
impact of these residual sources on derived parameters (Paper 4) the cosmic
parameter analysis
has been repeated {\it after} subtracting the residual contributions shown 
Figure~\ref{sources} from the observed power spectrum. 
The changes in the fitted parameters are small, the largest
effect being a reduction of 0.05 in $n_s$, representing a change of 
about 0.5 s.d.

\subsection{Galactic Foregrounds}

The diffuse Galactic foregrounds are discussed in Paper II, where it is 
suggested 
that the total contribution from the three components of the Galactic
foreground (synchrotron, free-free and possibly spinning dust) amount to no
more than about $\Delta T=5~\mu$K at an angular scale of $1^{\circ}$ ($\ell
\sim 200$). 
Here we discuss their effect on the results presented in this paper.
The free-free and synchrotron components are relatively well-known
and are each expected to contribute about $\Delta T=1-2~\mu$K. The spinning
dust component is more uncertain, but may be the dominant component, perhaps
contributing up to $\Delta T \approx 5~\mu$K. This is based on a
dust-correlated component with a correlation coefficient of
$10~\mu$K/(MJy~sr$^{-1}$) at $100~\mu$m. This coefficient is still to be
clearly demonstrated observationally. The power spectra of the diffuse
foregrounds falls with increasing $\ell$ (for example, \citet{giardino-01})
 and is likely to be half these values at $\ell \sim
500$ .

Any Galactic contribution adds in quadrature with the CMB signal and
hence at the position of the first CMB peak ($\ell \sim 200$) which has
$\Delta T_{{\rm CMB}}\approx 75~\mu$K, $5~\mu$K of foreground signal will
increase the observed signal by $0.17~\mu$K. Similarly, at $\ell \sim 500$
where $\Delta T_{{\rm CMB}}\approx 45~\mu$K, the increase will be $0.28~\mu$K
at the most. We see therefore, for the VSA fields, and at a frequency of 34
GHz, the contibution from Galactic foregrounds is likely to be negligible
($< 1$ percent).
A similar conclusion, albeit  based on observations of different regions of sky,
was reached by Halverson et al. (2002) who found that Galactic emission
made a negligible contribution to the observed CMB power spectrum.
 A more complete
cross-correlation analysis to investigate the contribution from
dust-correlated emission is in progress (Dickinson et al.  in prep.).

\subsection{Noise Estimation}

The likelihood analysis used for power-spectrum estimation requires an
accurate estimate of the rms noise level.  Since on individual baselines the
contribution of the CMB to the individual data samples is very small, the
noise level can be obtained directly from the standard deviation of the data,
after the filtering and flagging processes have been completed.  The noise
level associated with each of the binned visibilities (Section 3) is obtained
from an appropriately weighted combination of the noise levels of the data in
each bin.  As an additional check, the scatter in the data points contributing
to each bin has also been used to provide a noise estimate.  Consistent
estimates of the noise level were obtained by the two methods.  The overall
noise estimate is accurate to 2.5\%.

The sensitivity of the likelihood analysis to errors in the noise level has
been tested by analysing the same dataset with different assumed levels of
noise; the corresponding changes in estimated CMB power produced by
the above uncertainty in estimated noise level are less than 1\% for
$\ell$-values up to $\sim 750$.

\subsection{$\chi^{2}$ Tests}

As an additional check on the consistency of the data, we computed the
$\chi^{2}$ statistic for a variety of splits on the visibility data.
The data from each VSA field were split into two and the difference
vector, $\bmath{\Delta} = (\bmath{v}_{\rm{1}} - \bmath{v}_{\rm{2}})$,
formed where $\bmath{v}_{\rm{1}}$ and $\bmath{v}_{\rm{2}}$ are the
visibility vectors for each half of the data.  The $\chi^2$ statistic
was then calculated using
$\chi^2=\bmath{\Delta}^{\rm{t}}\bmath{C}_{\rm{N}}^{-1}\bmath{\Delta}$,
where $\bmath{C}_{\rm N}$ is the noise covariance matrix given by
$\bmath{C}_{\rm N}=(\bmath{C}_{\rm N1}+\bmath{C}_{\rm N2})$.  Since for
each field the sky signal measured in each of the two halves of the
data should be the same, the $\chi^2$ statistic can be used to test
for the presence of systematic errors in the data.  The data for each
VSA field were split in two according to observing epoch and the
$\chi^2$ values and associated significances are given in
Table\ref{chi}.  The data split on the VSA3 field also corresponds to
a split between day and night observations.  This was the only field
to be observed in this way, and its $\chi^2$-value confirms the
consistency of daytime versus nightime observation.

The consistency of the power spectra derived from each of the 3 VSA
mosaiced fields (VSA1M,VSA2M and VSA3M) was also compared by forming
the $\chi^2$ statistic on pairs of power spectra. In this case the
$\chi^2$ value is given by
\[
\chi^{2}=(\bmath{a}_{\rm1}-\bmath{a}_{\rm2})^{t}(\bmath{H}_{\rm1}^{-1}+\bmath{H}_{\rm2}^{-1})^{-1}(\bmath{a}_{\rm1}-\bmath{a}_{\rm2}),
\label{chi_eq}
\]
where $\bmath{a}_{1}$ and $\bmath{a}_{2}$ are the two sets of
bandpowers and $\bmath{H}_{\rm1}^{-1}$ and $\bmath{H}_{\rm2}^{-1}$ are
the corresponding inverse Hessian matrices.  Since $\chi^{2}$
statistic assumes that the likelihoods are Gaussian, we use the
Hessians calculated using an offset log-normal approximation (see
Paper IV for further details). The $\chi^{2}$ values (and
significances) for the VSA1M/VSA2M, VSA1M/VSA3M and VSA2M/VSA3M power
spectra comparisons are 7.6 (0.67), 9.82 (0.46) and 5.03 (0.89)
respectively. In each case there are ten degrees of freedom in the
power spectrum analysis.

\begin{table}
\begin{center}
\begin{tabular}{|c|c|c|c|}
\hline
Field & DOF & $\chi^2$ & Significance\\
\hline
VSA1     &977  & 1033.7 & 0.10\\ 
VSA1A    &1884 & 1947.7 & 0.15\\
VSA1B    &1387 & 1410.3 & 0.33\\
VSA2     & 915 & 984.2  & 0.06\\
VSA2-OFF &1384 & 1420.2 & 0.24\\
VSA3     &1287 & 1356.2 & 0.09\\
VSA3A    &2003 & 2094.2 & 0.08\\
VSA3B    &1584 & 1660.1 & 0.09\\
\hline
\end{tabular}
\caption{The $\chi^2$ values for data splits on each of the
  VSA fields.  In each case the visibility data from each field was
  split in two according to epoch and the $\chi^2$ of the difference
  vectors formed.  Also tabulated are the number of degrees of freedom
  (DOF) and the significance of each $\chi^2$ value; the significance is
  given as the probability to exceed the observed value in the
  $\chi^2$ cumulative distribution function.
\label{chi}}
\end{center}
\end{table}

\begin{table}
\begin{center}
\begin{tabular}{|c|c|c|c|}
\hline
$B$ & $\ell$ & \multicolumn{2}{c}{$\ell(\ell+1)C_{\ell}/2\pi [\mu {\rm K}^2]$} \\
\hline
1  & 142 & $3953_{-1248}^{+1709}$ &\\ 
1A & 184 &   & $5246_{-1211}^{+1493}$ \\ 
2  & 224 & $6200_{-1122}^{+1382}$ &\\ 
2A & 266 &   & $6494_{-1040}^{+1233}$\\
3  & 307 & $3496_{-661}^{+713 }$&\\
3A & 349 &   & $2080_{-416}^{+460}$\\
4  & 390 & $2122_{-416}^{+416 }$&\\
4A & 432 &   & $1954_{-371}^{+497}$\\
5  & 473 & $1498_{-460}^{+497}$ &\\
5A & 515 &   & $2452_{-535}^{+624}$\\
6  & 556 & $3246_{-705}^{+787}$ &\\
6A & 598 &   & $1998_{-705}^{+750}$\\
7  & 639 & $1207_{-662}^{+795}$ &\\
7A & 681 &   & $2162_{-787}^{+876}$\\
8  & 722 & $2039_{-869}^{+1003}$ &\\
8A & 764 &   & $666_{-665}^{+917}$\\
9  & 806 & $499_{-499}^{+1292}$ &\\
9A & 847 &   & $1954_{-1413}^{+1664}$\\
10 & 888 & $1914_{-1543}^{+1873}$ &\\
10A& 930 &   & $541_{-541}^{+2832}$ \\

\hline
\end{tabular}
\caption{\label{powerspec}The power spectrum from combining the three VSA
fields. The two sets of bin numbers (1, 1A etc) refer to the main and
alternate binnings, the latter being shifted by half a bin width. All the bins
have $\Delta \ell = 83$. The reported errorbars correspond to 68\% confidence
limits and were calculated by enclosing 68\% area under a likelihood curve
assuming independent bins. For further data analysis it is neccessary to use
full window functions and covariance matrices that can be downloaded from our
website.  In addition to these errors, there is an overall 7\% calibration
uncertainty in power.}
\end{center}
\end{table}

\begin{table}
\begin{tabular}{|c|c|c|c|c|c|c|}
\hline
$B$ & ${\cal C}_{B,B-2}$ & ${\cal C}_{B,B-1}$ & ${\cal C}_{B,B}$ &
${\cal C}_{B,B+1}$ & ${\cal C}_{B,B+2}$ & $\mbox{Cov}_{B,B} $\\

\hline

$1$    &            &            & $   1.00$  & $  -0.06$  & $   0.00$   &  $18.78$\\
$2   $ &            & $  -0.06 $ & $   1.00$  & $  -0.13$  & $   0.00$   &  $15.53$\\
$3   $ & $ 0.00  $  & $  -0.13 $ & $   1.00 $ & $  -0.07 $ & $   0.02 $ & $   4.86$\\
$4   $ & $ 0.00   $ & $  -0.07 $ & $   1.00 $ & $  -0.18 $ & $   0.06 $ & $   1.98$\\
$5   $ & $ 0.02   $ & $  -0.18 $ & $   1.00 $ & $  -0.18 $ & $   0.01 $ & $   2.46$\\
$6   $ & $ 0.06   $ & $  -0.18 $ & $   1.00 $ & $  -0.15 $ & $   0.05 $ & $   6.19$\\
$7   $ & $ 0.01   $ & $  -0.15 $ & $   1.00 $ & $  -0.22 $ & $   0.02 $ & $   5.96$\\
$8   $ & $ 0.05   $ & $  -0.22 $ & $   1.00 $ & $  -0.16 $ & $   0.03 $ & $  10.06$\\
$9   $ & $ 0.02   $ & $  -0.16 $ & $   1.00 $ & $  -0.21 $ & $    $     & $  12.41$\\
$10  $ & $ 0.03   $ & $  -0.21 $ & $   1.00 $ & $    $     & $    $     & $  35.88$\\

\hline
\end{tabular}
\caption{The correlation matrix ${\cal C}_{i,j}$ for the combined VSA data set
(main binning only). These are calculated by normalising the covariance matrix
of errors. Note that the correlation is only significant for adjacent bins.
The values of matrix for which ${\cal C}_{i,j}$ is not reported can be assumed
to be zero. The final column gives the diagonal elements of the covariance
matrix in units of $10^5 \times \mu \rm{K}^4 $
\label{covar}}
\end{table}

\section{The Power Spectrum}

\begin{figure*}
\centering
\epsfig{file=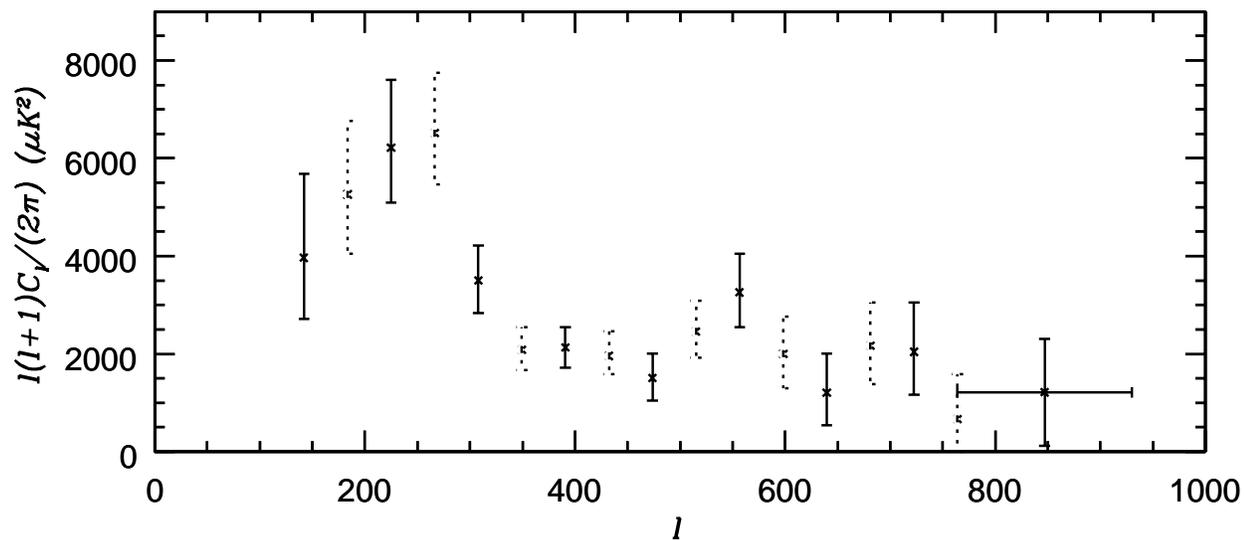,width=170mm}
\caption{Combined CMB power spectrum from the three mosaiced
     VSA fields.  The error-bars represent $1\sigma$ limits; the two sets of
      data points correspond to alternative interleaved binnings of the data. }
\label{CMB_doublebin}
\end{figure*}

The filtered and source-subtracted data for each of the VSA fields have been
analysed using the {\sc Madcow} software package \citep{klaus} as described in
Section 3.  The combination of the three mosaiced spectra is shown in
Figure~\ref{CMB_doublebin}. The bin width used is $\Delta \ell = 83$, which
gives weakly-correlated errors in each bin. The actual correlations between
bins are given by the correlation matrix (Table~\ref{covar}). To reduce the
bias in assessing features in the power spectrum caused by the settings of the
bin centres, we have also calculated the power spectrum with bin centres
shifted by one half a bin width to the right of the original bin centres.
These results are shown in Figure~\ref{CMB_doublebin} with dashed error bars.
Adjacent `double-binned' points are highly correlated but do sensibly sample
the power spectrum of our data.  Numerical values for both binnings are given
in Table~\ref{powerspec}.

The plotted error bars contain the contributions from both thermal noise and
sample variance, but not calibration errors, which introduce a completely
correlated uncertainty in all the points of $\pm 7$ percent. Errors from
pointing and primary beam uncertainties are negligible. Since the temperature
sensitivity of the VSA compact configuration falls off dramatically after
$\ell=800$, all data above this have been binned together.

\section{Discussion}

\subsection{The VSA power spectrum}

The power spectrum shown in Figure~\ref{CMB_doublebin} shows a clear detection
of the first peak at $\ell \simeq 220$, and power at the level of about $2000
\, \mu$K$^2$ between $\ell = 300$ and $\ell = 900$. We have attempted to
quantify the detection of a second peak at $\ell \simeq 550$, as this is the
region of the power spectrum with the largest anti-correlations between
adjacent bins (see Table~\ref{covar}); bin 6 centred at $\ell = 556$ is
anti-correlated with its neighbours at the $\simeq 20$~percent level.  We made
Monte Carlo simulations of the five $C_{\ell}$ points between $\ell = 390$ and
$\ell = 722$. Points were draw from the 5-dimensional Gaussian distribution
described by the actual correlation matrix of these points, but with mean
values equal to the weighted mean of the five actual points. Following
\cite{hobson-96}, who define a normalised convexity about a power spectrum
point, we calculated the change in normalised convexity of the power spectrum
about the inner 3 points, defined by
\[
\Sigma={\cal C \over \sigma},
\]
where ${{\cal C}} = (C_4+C_8 -3(C_5 + C_7))/2 + 2C_6$ and $\sigma$ is the
overall error in ${\cal C}$ given from the errors in the individual points
$\sigma_i$ by $\sigma^2 = (\sigma^2_4 + \sigma^2_8 + 9(\sigma^2_5 +
\sigma^2_7))/4 + 4\sigma^2_6$. This effectively compares the hypothesis that
the power spectrum in this region is flat, to the one that it is described by
`trough--peak--trough'.

In 1000 realisations, we found only 27 instances of $\Sigma$ being larger than
the value observed in the real data of $\Sigma = 2.2$. We therefore conclude
that the observed second peak is detected at $97$ percent confidence. The
power spectrum is completely consistent with the adiabatic inflationary
models, fits to which are discussed in Paper IV.

\subsection{Comparison with other experiments}

\begin{figure*}
 \centering
  \epsfig{file=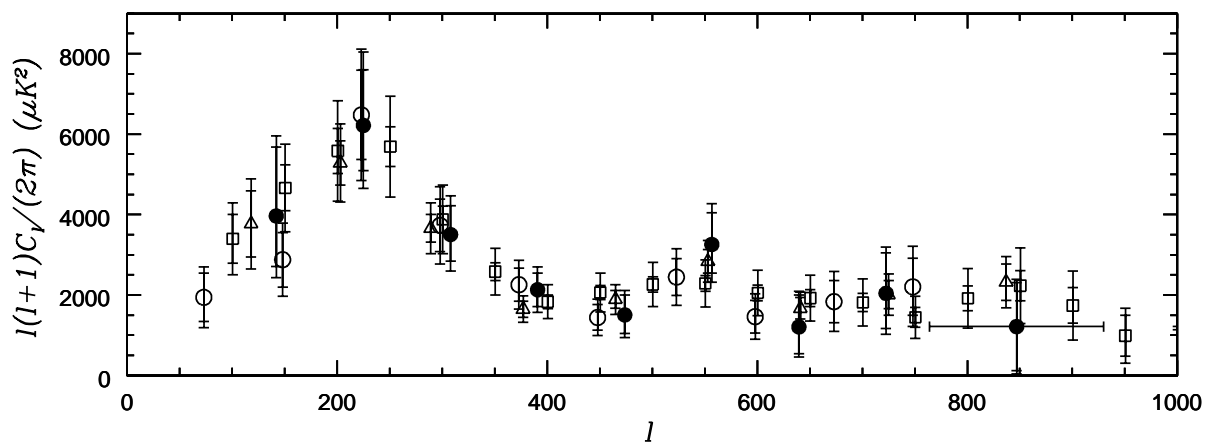,height=170mm,angle=270}
\caption{A comparison of the VSA data (filled circles) with results from 
  the BOOMERANG (open squares), MAXIMA (open circles) and DASI (open
  triangles) experiments. Two sets of error bars are plotted for each
  data set; the smaller of the two indicate only random errors, whilst
  the larger bars indicate the amount by which the inner points could move
  due to absolute calibration and beam uncertainty. In each case the
  error bars indicate $1\sigma$ limits.
  \label{ps-comparison}}
\end{figure*}

In Figure~\ref{ps-comparison} we compare the new VSA power spectrum plotted
with those from BOOMERANG~\citep{netterfield-01},
DASI~\citep{halverson-02},~MAXIMA \citep{lee-01}. Only single-binned
(weakly-correlated) points are shown. We have attempted to compare the random
and correlated errors on the various experiments in a consistent way,
difficult though this is on a single plot. Two sets of error bars are shown
for each plot; the smaller bars indicate 68~percent confidence limits from the
random (thermal and sample variance) errors, while the larger error bars
represent systematic (calibration and beam) errors as reported -- {\em all}
the points from a single experiment are able to move up or down within the
larger error bars.  It is significant that, although the overall errors of the
different experiments are comparable, the relative contribution of systematic
errors (including beam uncertainty) is much smaller for the interferometric
data.

The agreement between the experiments on the existence, heights and positions
of two peaks and of power at higher $\ell$, is evident. This is particularly
significant given the very different experimental techniques involved and the
different foregrounds and systematic errors faced by the different
experiments. The points in Figure~\ref{ps-comparison} have been obtained over
a frequency range of $26$--$150$~GHz, and by ground-based interferometers and
balloon-borne scanned total-power telescopes. They are all from different
regions of the sky, the calibrations are all independent and based on
different absolute calibration sources, and for the two low-frequency
experiments foreground sources have been subtracted in different ways, yet the
agreement of the power spectra is striking.

A detailed comparison between the experiments is difficult to do from the data
points alone because of the correlated errors between points for each
experiment. To make a meaningful comparison it is necessary to fit the
underlying power spectrum to each data set, taking into account the
correlations, and to compare the parameters that describe the power
spectrum. In principle many parameterisations of the power spectrum would
suffice for this comparison, but in practice it is obviously sensible to use
the standard adiabatic cold dark matter power spectrum models, and fits to
these models for the VSA and other experiments are considered in detail in
Paper IV.

\section{Conclusions}

We have derived the power spectrum of the CMB anisotropies from the first
year's VSA observations, made using its compact array configuration. We
measure the flat band power in 10 weakly-correlated  bins of width $\Delta \ell
= 83$ between $\ell \simeq 150 $ and $\ell \simeq 900 $. The results are subject to a
calibration uncertainty of $\pm 7$ percent in power, with negligible beam
uncertainty. The contribution to the power spectrum from diffuse Galactic
emission and residual radio sources is also negligible. Our results are in
excellent agreement with other recent measurements as regards the amplitude
and position of two peaks in the power spectrum; power is also detected out to
the resolution limit of the experiment at $\ell \simeq 900$.

Band powers, correlation matrices and window functions are available from

\noindent \verb+http://mrao.cam.ac.uk/telescopes/vsa/results.html+.

\section*{ACKNOWLEDGEMENTS}

We thank the staff of the Mullard Radio Astronomy Observatory, Jodrell Bank
Observatory and the Teide Observatory for invaluable assistance in the
commissioning and operation of the VSA. The VSA is supported by PPARC and the
IAC. Partial financial support was provided by Spanish Ministry of Science and
Technology project AYA2001-1657.  A. Taylor, R. Savage, B. Rusholme,
C. Dickinson acknowledge support by PPARC studentships. K. Cleary and
J. A. Rubi\~no-Martin acknowledge Marie Curie Fellowships of the European
Community programme EARASTARGAL, ``The Evolution of Stars and Galaxies'',
under contract HPMT-CT-2000-00132. K. Maisinger acknowledges support from an
EU Marie Curie Fellowship. A. Slosar acknowledges the support of St. Johns
College, Cambridge. We thank Professor Jasper Wall for assistance and advice
throughout the project.

\clearpage

\clearpage

\label{lastpage}
\bibliography{cmb_refs}
\bibliographystyle{mn2e}
\bsp
\end{document}